%% file: synmut.tex
\documentclass[rmp, twocolumn]{revtex4}
\usepackage[english]{babel}
\usepackage[utf8x]{inputenc}
\usepackage{amsmath,amsfonts,amssymb,eucal,eurosym,textcomp}
\usepackage{color}
\usepackage{graphicx}
\usepackage[caption=false]{subfig}
\usepackage{natbib}
\usepackage{pslatex}
\usepackage[colorlinks,linkcolor=red,citecolor=blue]{hyperref}
\graphicspath{{./figures/}}

\newcommand{\pfix}{P_{\mathrm{fix}}}

\newcommand{\rev}{\textit{rev}}
\newcommand{\FIG}[1]{Fig.~\ref{fig:#1}}

\newcommand{\env}{\textit{env}}
\newcommand{\shankaregion}{C2-V5}
\newcommand{\PCApat}{1}
\newcommand{\timedependence}{3}
\newcommand{\withinepi}{4}

\newcommand{\Author}{Fabio~Zanini and Richard~A.~Neher}
\newcommand{\Title}{Deleterious synonymous mutations hitchhike to high frequency in HIV-1 \env~evolution}
\newcommand{\Affiliation}{Max Planck Institute for Developmental Biology, 72076 T\"ubingen, Germany}
\newcommand{\Keywords}{{HIV}, {synonymous}, {population genetics}}
\hypersetup{pdfauthor={\Author}, pdftitle={\Title}, pdfkeywords={\Keywords}}
\begin{document}
\title{\Title}
\author{\Author}
\affiliation{\Affiliation}
\date{\today}

\begin{abstract}
\noindent
Intrapatient HIV-1 evolution is dominated by selection on the protein level in the
arms race with the adaptive immune system. When cytotoxic CD8${}^+$ T-cells or 
neutralizing antibodies
target a new epitope, the virus often escapes via nonsynonymous mutations that
impair recognition. Synonymous mutations do not affect this interplay and are
often assumed to be neutral.
We analyze longitudinal intrapatient data from the \shankaregion{} part of the
envelope gene (\env{}) and observe that synonymous derived alleles rarely
fix even though they often reach high frequencies in the viral population.
We find that synonymous mutations that disrupt base pairs in RNA stems flanking
the variable loops of gp120 are more likely to be lost than other synonymous
changes, hinting at a direct fitness effect of these stem-loop structures in the
HIV-1 RNA.
Computational modeling indicates that these synonymous mutations have a
(Malthusian) selection coefficient of the order of $-0.002$ and that they are
brought up to high frequency by hitchhiking on neighboring beneficial
nonsynonymous alleles.
The patterns of fixation of nonsynonymous mutations estimated from the
longitudinal data and comparisons with computer models
suggest that escape mutations in \shankaregion{} are only transiently
beneficial, either because the immune system is catching up or because of
competition between equivalent escapes.

\end{abstract}
\maketitle
\section{Introduction}
HIV-1 evolves rapidly within a single host during the course of the infection.
This evolution is driven by strong selection imposed by the host immune system
via cytotoxic CD8${}^+$ T-cells (CTLs) and neutralizing antibodies
(nAbs)~\citep{rambaut_causes_2004} and facilitated by the high mutation rate
~\citep{mansky_lower_1995,abram_nature_2010}. When the host develops a CTL or
nAb response against a particular HIV-1 epitope, mutations in the viral genome that
reduce or prevent recognition of the epitope frequently emerge. Escape mutations
in epitopes targeted by CTLs typically evolve during early infection and spread
rapidly through the population~\citep{mcmichael_immune_2009}. During chronic
infection, the most rapidly evolving parts of the HIV-1 genome are the variable
loops V1-V5 in the envelope protein gp120, which change to avoid recognition by
nAbs. Escape mutations in \env, the gene encoding gp120, spread through the
population within a few months.
Consistent with this time scale, it is found that serum from a particular time
typically neutralizes virus extracted more than 3-6 months earlier but not contemporary
virus \citep{richman_rapid_2003}.

Escape mutations are selected because they change the amino acid sequence
of viral proteins in a way that reduces antibody binding or epitope presentation. 
Conversely, synonymous mutations are commonly used as
approximately neutral markers in studies of viral evolution. Neutral markers are
very useful since their dynamics can be compared to that of putatively
functional sites to detect purifying or directional selection
\citep{Bhatt:2011p43255,Hurst:2002p32608,Chen:2004p22606}. In addition to
maintaining protein function and avoiding the adaptive immune recognition,
however, the HIV-1 genome has to ensure efficient processing and translation,
nuclear export, and packaging into the viral capsid: all these processes operate
at the RNA level and are sensitive to synonymous changes since these processes
often depend on RNA folding. For example, the HIV-1 \rev{} response element (RRE)
in \env{} enhances nuclear export of full length or partially spliced viral
transcripts via a complex stem-loop RNA structure~\citep{fernandes_hiv-1_2012}.
Another well studied case is the interaction between viral reverse
transcriptase, viral ssRNA, and the host tRNA$^\text{Lys3}$: the latter is
required for priming reverse transcription (RT) and is bound by a pseudoknotted
RNA structure in the viral 5' untranslated region~\citep{barat_interaction_1991,
paillart_vitro_2002}.

Even in the absence of important RNA structures, synonymous codons do not evolve
completely neutrally. Some codons are favored over others in many species
\citep{plotkin_synonymous_2011}. Recent studies have shown that genetically
engineered HIV-1 strains with altered codon usage can in some cases produce more
viral protein, but in general replicate less efficiently
\citep{ngumbela_quantitative_2008, li_codon-usage-based_2012,
keating_rich_2009}. Codon deoptimization has been suggested as an attenuation
strategy for polio and influenza~\citep{mueller_live_2010,coleman_virus_2008}.
Purifying selection beyond the protein sequence is therefore expected
\citep{forsdyke_reciprocal_1995,snoeck_mapping_2011}, and it has been shown that
rates of evolution at synonymous sites vary along the HIV-1 genome
\citep{mayrose_towards_2007}. Positive selection through the host adaptive
immune system, however, is restricted to changes in the amino acid sequence.

In this paper, we characterize the dynamics of synonymous mutations in \env{}
and show that a substantial fraction of these mutations is deleterious.  We
argue that synonymous mutations reach high frequencies via genetic hitchhiking
due to limited recombination in HIV-1 populations~\citep{neher_recombination_2010,
batorsky_estimate_2011}. We then compare our observations to computational
models of HIV-1 evolution and derive estimates for the effect of synonymous mutations
 on fitness.  Extending the analysis of fixation probabilities to the
nonsynonymous mutations, we show that time-dependent selection or strong
competition of escape mutations inside the same epitope are necessary to explain
the observed patterns of fixation and loss.

\section{Results}
\begin{figure}
\begin{center}
\subfloat{\includegraphics[width=\linewidth]
{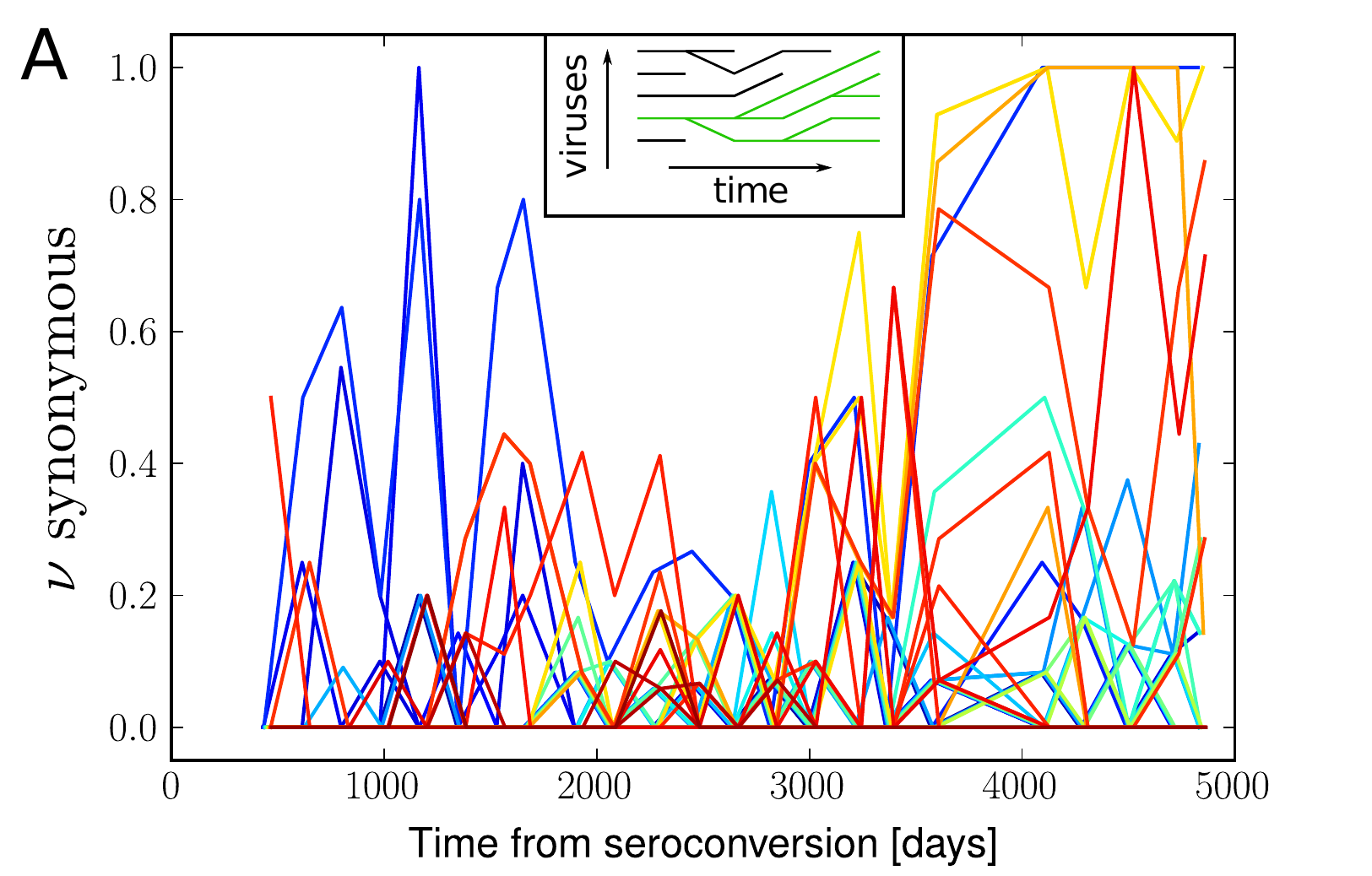}
\label{fig:aftsyn}}\\
 \subfloat{\includegraphics[width=\linewidth]
{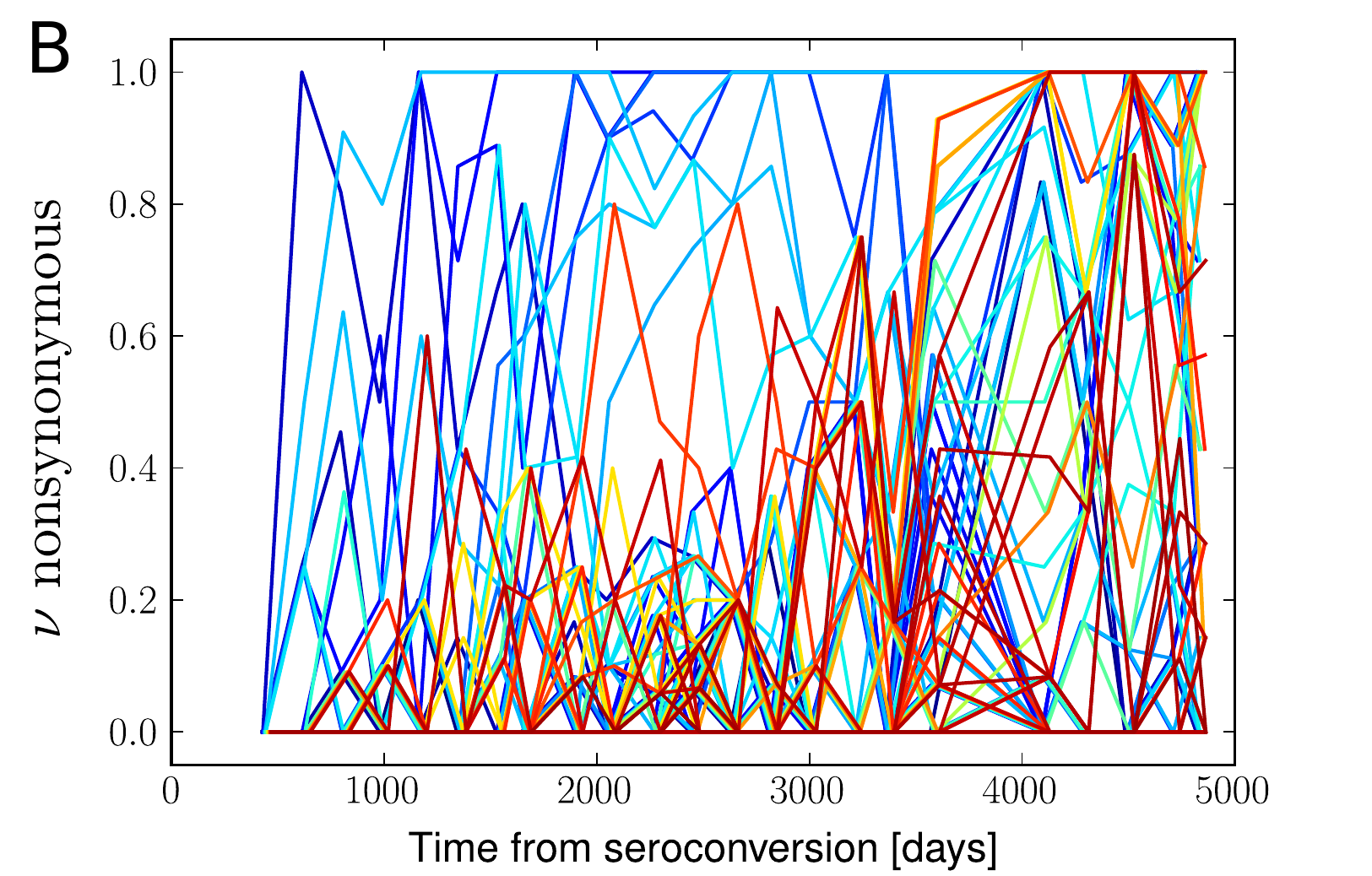}
\label{fig:aftnonsyn}}
\caption{Time series of frequencies
of synonymous (A) and nonsynonymous (B) derived alleles in \env, 
\shankaregion, from patient 10~\cite{shankarappa_consistent_1999}.
While many nonsynonymous mutations  fix, few synonymous
mutations do even though they are frequently observed at intermediate
frequencies. Colors indicate the position of the site along the \shankaregion{} region
(blue to red). Inset: the fixation probability $\pfix$ of a neutral
mutation is simply the likelihood that the future common ancestor at this 
position is currently carrying it, i.e. the mutation frequency $\nu$.}
\label{fig:aft}
\end{center}
\end{figure}

The central quantity we investigate is the probability of fixation of a
mutation, conditional on its population frequency. A neutral mutation
segregating at frequency $\nu$ has a probability $\pfix(\nu) = \nu$ to
spread through the population and fix; in the rest of the cases, i.e. with
probability $1-\nu$, it goes extinct. As illustrated in the inset of
\FIG{aftsyn}, this is a  consequence of the fact that (i) exactly one
individual in the current population will be the common ancestor of the entire
future population at a particular site and (ii) this ancestor has a probability
$\nu$ of carrying the mutation (assuming the neutral mutation is not
preferentially associated with genomes of high or low fitness). Deleterious or
beneficial mutations fix less or more often than neutral ones, respectively.
\FIG{aft} shows the time course of the frequencies of all synonymous and
nonsynonymous mutations observed in \env, \shankaregion, in patient
p10~\citep{shankarappa_consistent_1999}. Despite many synonymous
mutations reaching high frequency, few fix (panel~\ref{fig:aftsyn}); in
constrast, many nonsynonymous mutations fix (panel~\ref{fig:aftnonsyn}).
Strictly speaking, no mutation in the HIV-1 population ever fixes because the 
mutation rate and the population size are large. Therefore, we define ``fixation'' or ``loss''
by not observing the mutation in the sample.

\subsection{Synonymous polymorphisms in \env, \shankaregion, are mostly deleterious}
\begin{figure}
\begin{center}
\subfloat{\includegraphics[width=0.9\linewidth]{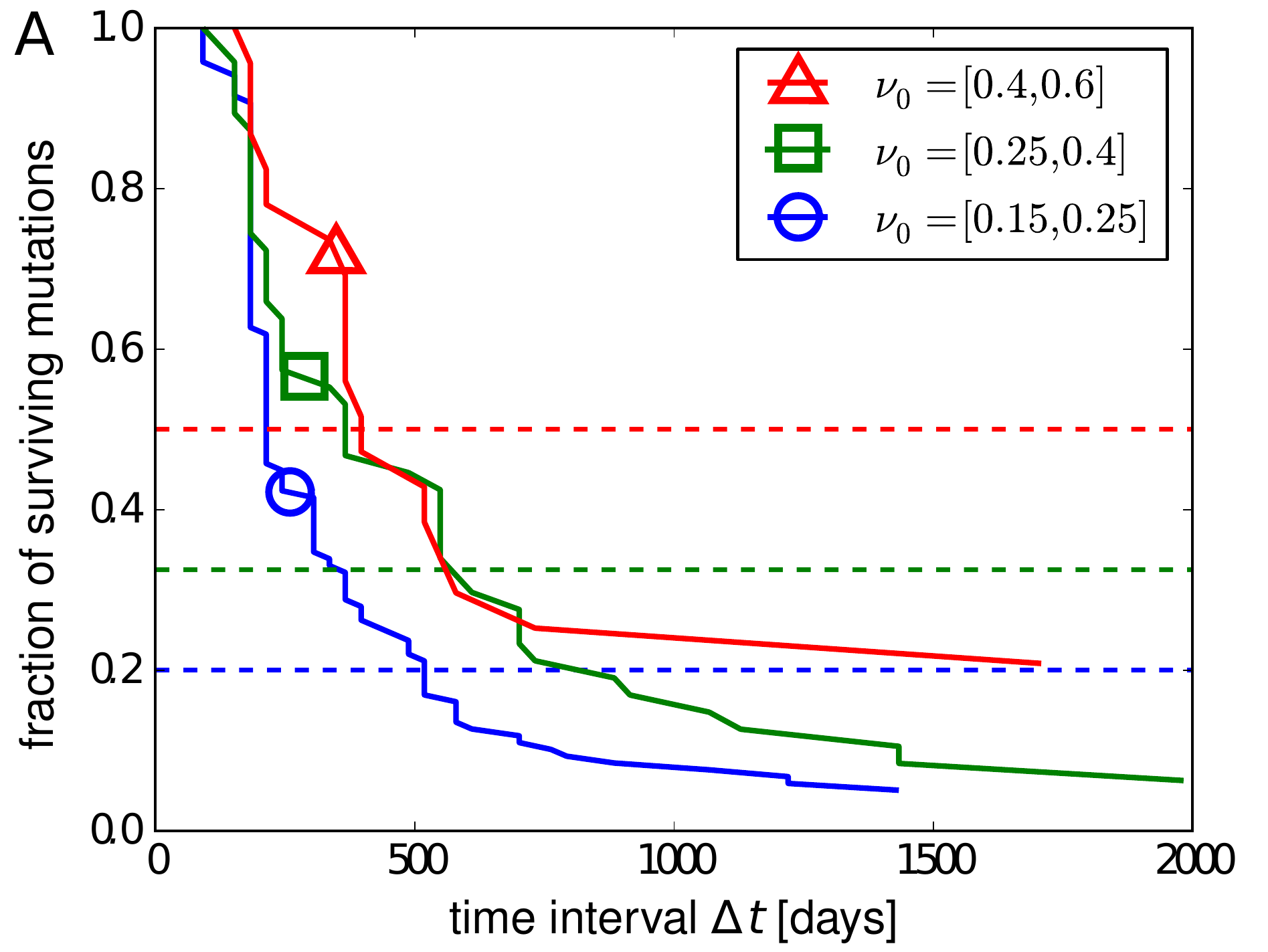}
\label{fig:fixp1}}\\
\subfloat{\includegraphics[width=0.9\linewidth]{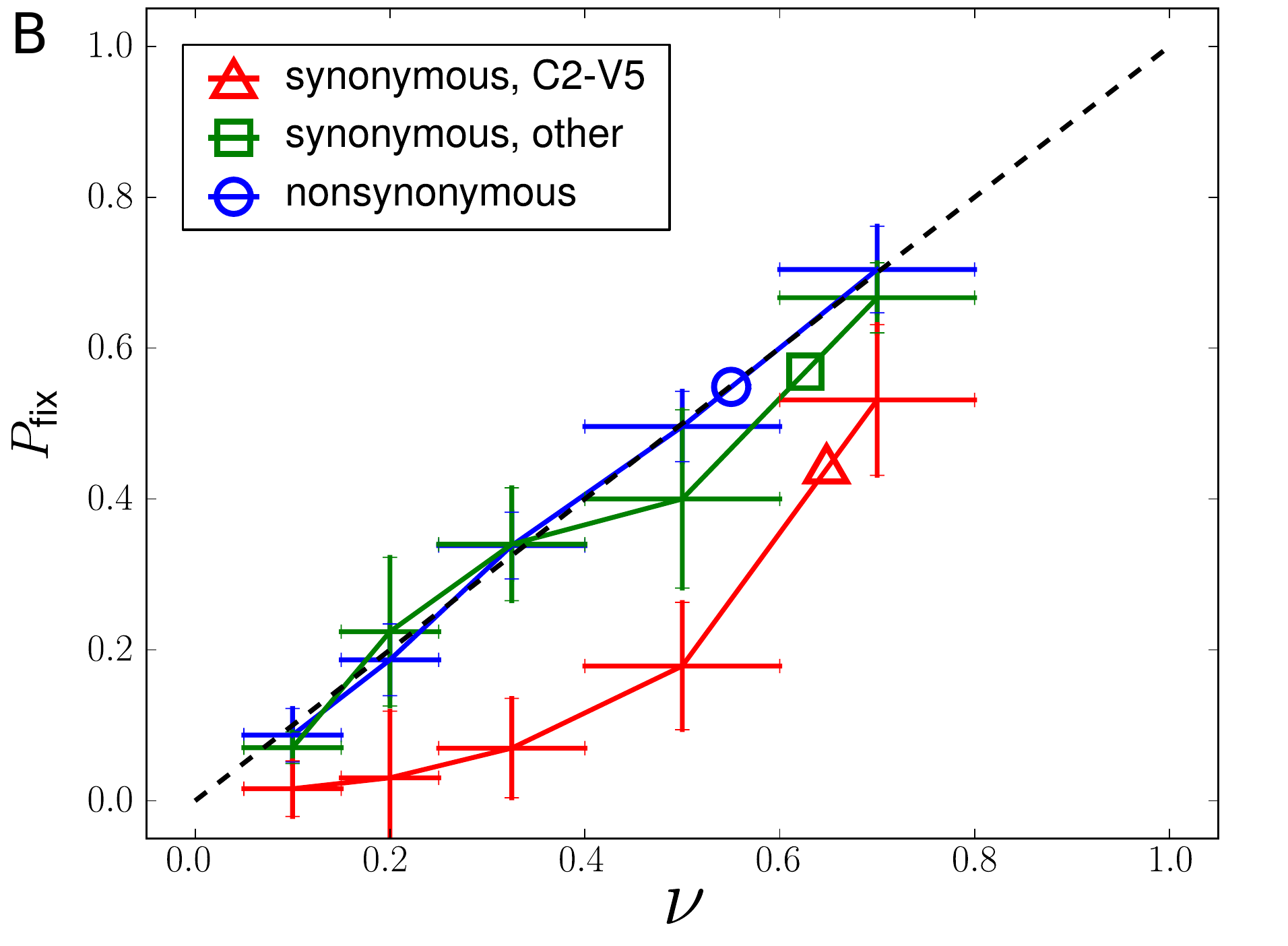}
\label{fig:fixp2}}
\caption{Fixation and loss of synonymous mutations.
Panel A) shows how quickly synonymous mutations are purged from the populations. 
Specifically, the figure shows the fraction of mutations that are still observed
after $\Delta t$ days, conditional on being observed in one of the three frequency 
intervals (different colors). 
In each frequency interval, the fraction of synonymous
mutations that ultimately survive is the fixation probability $\pfix$ conditional on the
initial frequency. The neutral expectation for $\pfix=\nu_0$ is indicated by 
dashed horizontal lines.
Panel B) shows the fixation probability of derived synonymous
alleles as a function of $\nu_0$. Polymorphisms within \shankaregion{} fix less
often than expected for neutral mutations (indicated by the diagonal line).
This suppression is not observed in other parts of \env~or for nonsynonymous mutations.
The horizontal error bars on the abscissa are bin sizes, the vertical ones the
standard deviation after 100 patient bootstraps of the data. Data from
refs.~\cite{shankarappa_consistent_1999,liu_selection_2006, bunnik_autologous_2008}.}
\label{fig:fixp}
\end{center}
\end{figure}

We study the dynamics and fate of synonymous mutations more quantitatively by
analyzing data from seven patients from
\citet{shankarappa_consistent_1999} and \citep{liu_selection_2006} as well as three patients from
\citet{bunnik_autologous_2008} (patients whose viral population was structured
were excluded from the analysis; see methods and \figurename~S\PCApat).  The
former data set is restricted to the \shankaregion{} region of \env, while
the data from \citet{bunnik_autologous_2008} cover the majority of \env.
Considering all mutations in a frequency interval $[\nu_0-\delta\nu, \nu_0+\delta\nu]$ at some time
$t$, we calculate the fraction that are still observed at later times $t+\Delta t$. Plotting this fraction against the time interval $\Delta t$, we see that
most synonymous mutations segregate for roughly one year and are lost much more
frequently than expected (panel \ref{fig:fixp1}). The long-time probability of
fixation, $\pfix$, is shown as a function of the
initial frequency $\nu_0$ in panel~\ref{fig:fixp2} (red line). 
We find that $\pfix$ of synonymous mutations is far below
the neutral expectation in  \shankaregion.  Outside of \shankaregion, using data from
\citet{bunnik_autologous_2008} only, we find no such reduction in $\pfix$.
Restricted to the \shankaregion{} region, the sequence samples from
\citet{bunnik_autologous_2008} are fully compatible with data from
\citet{shankarappa_consistent_1999}. The nonsynonymous mutations seem to follow
more or less the neutral expectation (blue line) -- a point to which we will
come back below.

When interpreting these results for the fixation probabilities, it is important
to distinguish between random mutations and polymorphisms observed at a certain
frequency since the latter have already been filtered by selection.
A polymorphism could be beneficial to the virus and on its way to fixation. In
this case, we expect that it fixes almost surely given that we see it at high
frequency. If, on the other hand, the polymorphism is deleterious it must have
reached a high frequency by chance (genetic drift or hitchhiking), and
we expect that selection drives it out of the population again. Hence our
observations suggest that many of the synonymous polymorphisms at intermediate
frequencies in the part of \env{} that includes \shankaregion{} are
deleterious, while outside this region most polymorphisms are roughly
neutral. Note that this does not imply that all synonymous mutations outside 
\shankaregion{} are neutral -- only those mutations observed at high frequencies, which
have experienced selection for some time, tend to be neutral.

\subsection{Synonymous mutations in \shankaregion{} tend to disrupt conserved RNA stems}
One possible explanation for lack of fixation of synonymous mutations in
\shankaregion{} is secondary structures in the viral RNA, the disruption of which
is deleterious to the virus \citep{forsdyke_reciprocal_1995,
snoeck_mapping_2011, sanjuan_interplay_2011}.

\begin{figure}
\begin{center}
\subfloat{\includegraphics[height=0.47\linewidth]{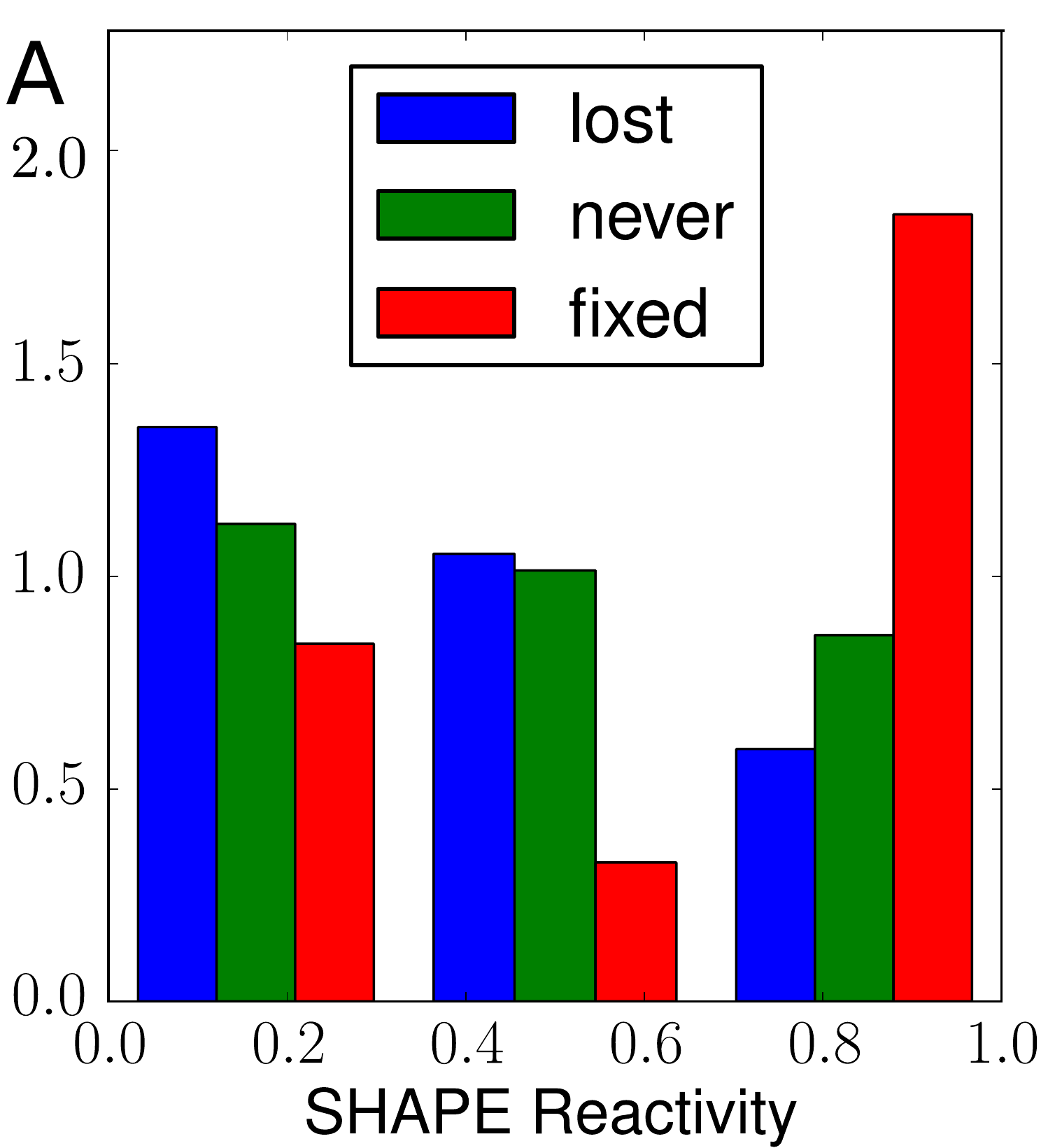}\label{fig:SHAPEA}}
\subfloat{\includegraphics[height=0.46\linewidth]{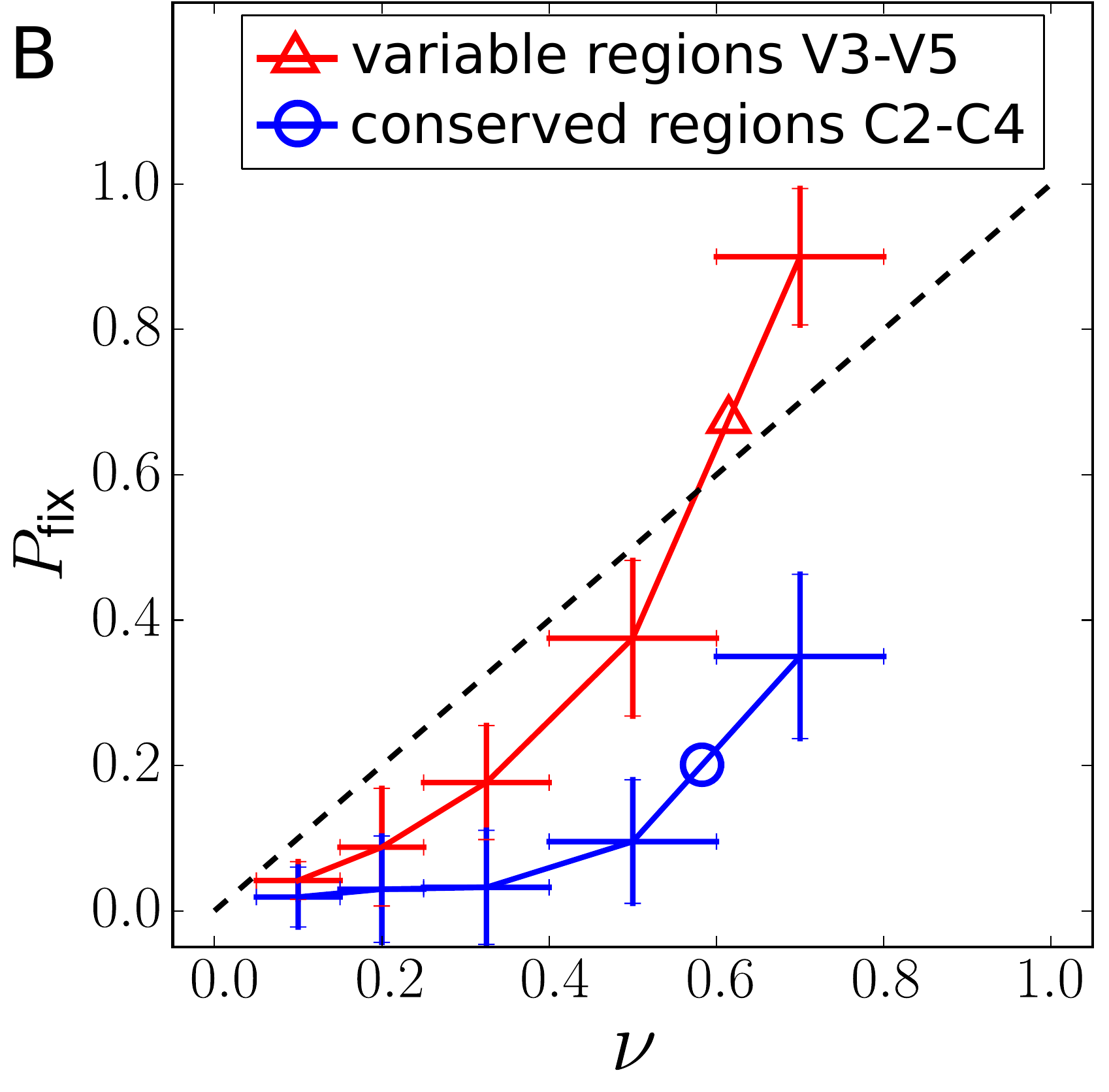}\label{fig:SHAPEB}}
\caption{Permissible synonymous mutations tend to be unpaired.
Panel A) shows the distribution of SHAPE reactivities among sites at which synonymous 
mutations fixed (red), sites at which mutations reached frequencies above 15\% but
were susequently lost (blue), and sites at which no mutations were observed (green) 
(all categories are restricted to the regions V1-V5$\pm 100$bp).
Sites at which mutations fixed tend to have higher SHAPE reactivities, corresponding to
less base pairing, than those at which mutations are lost.
Sites at which no mutations are observed show an intermediate distribution of SHAPE values.
Panel B) shows the fixation probability of synonymous mutations in
\shankaregion{} separately for variable regions V3-V5 and the connecting conserved 
regions C2-C4 that harbor RNA stems. As expected, the fixation probability is lower
inside the conserved regions. Data from Refs.~\cite{shankarappa_consistent_1999,
bunnik_autologous_2008, liu_selection_2006}.}
\label{fig:SHAPE}
\end{center}
\end{figure}

The propensity of nucleotides in the HIV-1 genome to form base pairs has been
measured using the SHAPE assay, a biochemical reaction preferentially altering
unpaired bases (the HIV-1 genome is a single stranded RNA)
\citep{watts_architecture_2009}. The SHAPE assay has shown that the variable
regions V1-V5 tend to be unpaired, while the conserved regions between those
variable regions form stems.  We aligned the within-patient sequence samples 
to the reference NL4-3 strain used by \citet{watts_architecture_2009} and 
thereby assigned SHAPE reactivities to most positions in the alignment. 
We then calculated the distributions of SHAPE reactivities for synonymous 
polymorphisms that fixed or were subsequently lost (only polymorphisms with 
frequencies above 15\%).
As shown in \FIG{SHAPEA}, the reactivities of fixed alleles (red
histogram) are systematically larger than those of alleles that are lost (blue)
(Kolmogorov-Smirnov test on the cumulative distribution, $p\approx 0.002$). In
other words, alleles that are likely to break RNA helices are also more likely
to revert and finally be lost from the population. The average over all
mutations that are not observed (green) lies between  those that fix and
those that get lost. Note that this analysis will be sensitive only at positions
where the base pairing pattern of NL4-3 agrees with that of each patient's
initial consensus sequence (it is thus statistically conservative).

To test the hypothesis that mutations in \shankaregion{} are lost because they
break stems in the conserved stretches between the variable loops, we consider
mutations in variable loops and conserved parts separately. The greatest
depression in fixation probability is observed in the conserved stems, while the
variable loops show little deviation from the neutral signature, see
\FIG{SHAPEB}. This is consistent with important stem structures in conserved
regions between loops.

In addition to RNA secondary structure, we have considered other possible
explanations for a fitness cost of some synonymous mutations, in particular codon
usage bias (CUB). HIV-1 is known to prefer A-rich codons over highly expressed
human codons~\citep{jenkins_extent_2003,kuyl_biased_2012}. We do not find,
however, any evidence for a contribution of average CUB to the ultimate fate of
synonymous alleles; consistently, HIV-1 does not seem to adapt its codon usage to
its human host cells at the macroevolutionary level \citep{kuyl_biased_2012}.

\subsection{Deleterious mutations are brought to high frequency by hitchhiking}

While the observation that some fraction of synonymous mutations is deleterious
is not unexpected, it seems odd that we observe them at high population
frequency and that the fixation probability is reduced only in parts of the
genome (in \shankaregion{} but not in the rest of \env{}; compare the red
triangle line versus the green square line in \FIG{fixp2}).
The region \shankaregion{} undergoes frequent adaptive changes to evade
recognition by neutralizing antibodies \cite{williamson_adaptation_2003,
richman_rapid_2003}. Due to the limited amount of recombination in HIV-1
\cite{neher_recombination_2010, batorsky_estimate_2011}, deleterious mutations
that are linked to adaptive variants can reach high frequency. This process is
known as hitchhiking \citep{smith_hitch-hiking_1974} or genetic draft
\citep{gillespie_genetic_2000,neher_genetic_2011}. Hitchhiking is  apparent in
\FIG{aft}, which shows that many mutations change rapidly in frequency as a
flock. 

The approximate magnitude of the deleterious effects can be estimated from
\FIG{fixp1}, which shows the distribution of times after which synonymous
alleles at intermediate frequencies become fixed or lost. The typical time to
loss is of the order of 500 days. If this loss is driven by the deleterious
effect of the mutation, this corresponds to deleterious effects $s_d$ of the
order of $- 0.002$ per day. (This is only an average estimate: every single
mutation is expected to have a slightly different fitness effect.)

\begin{figure}
\begin{center}
\subfloat{\includegraphics[width=0.9\linewidth]{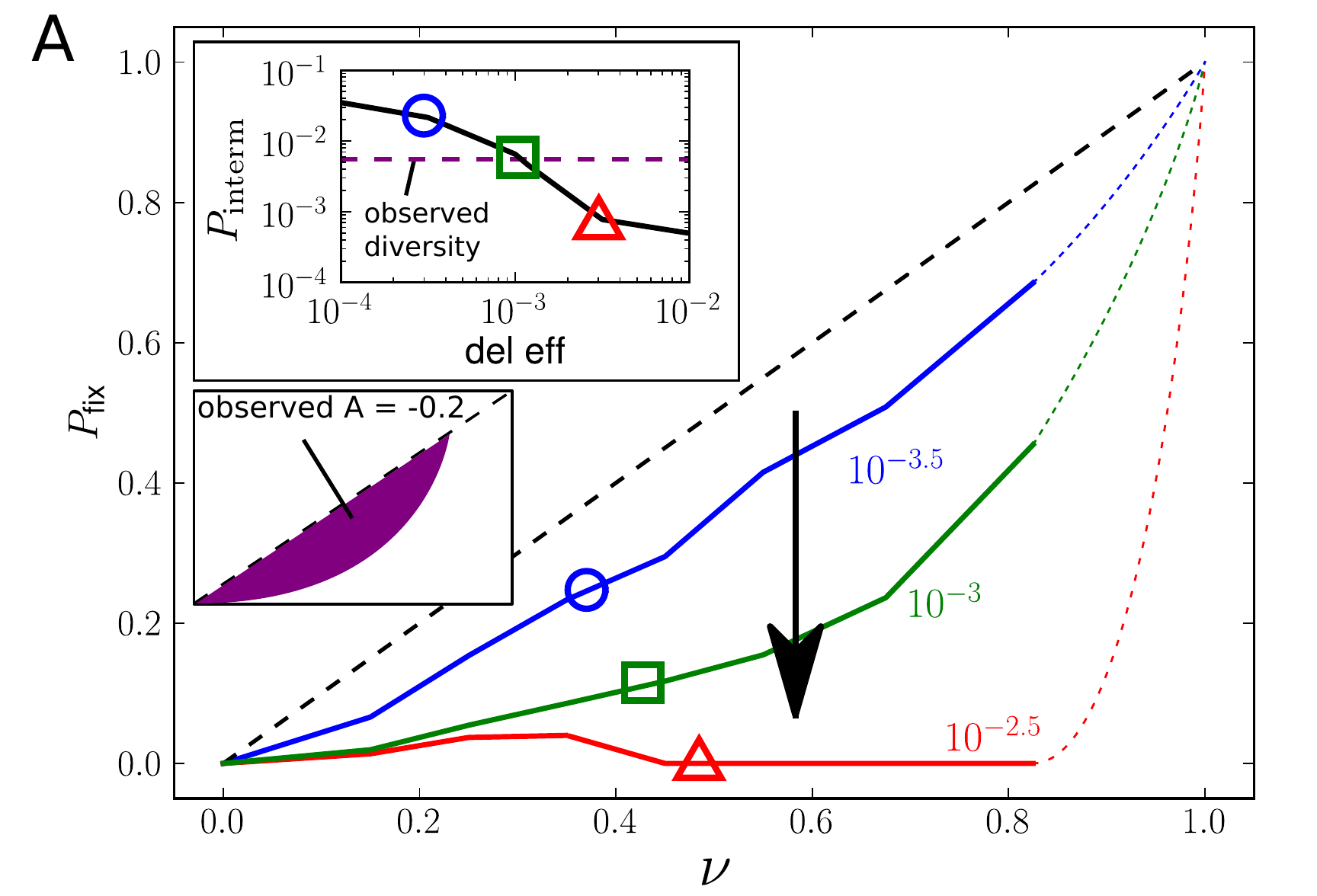}
\label{fig:simfixpvar}}\\
\subfloat{\includegraphics[width=0.9\linewidth]{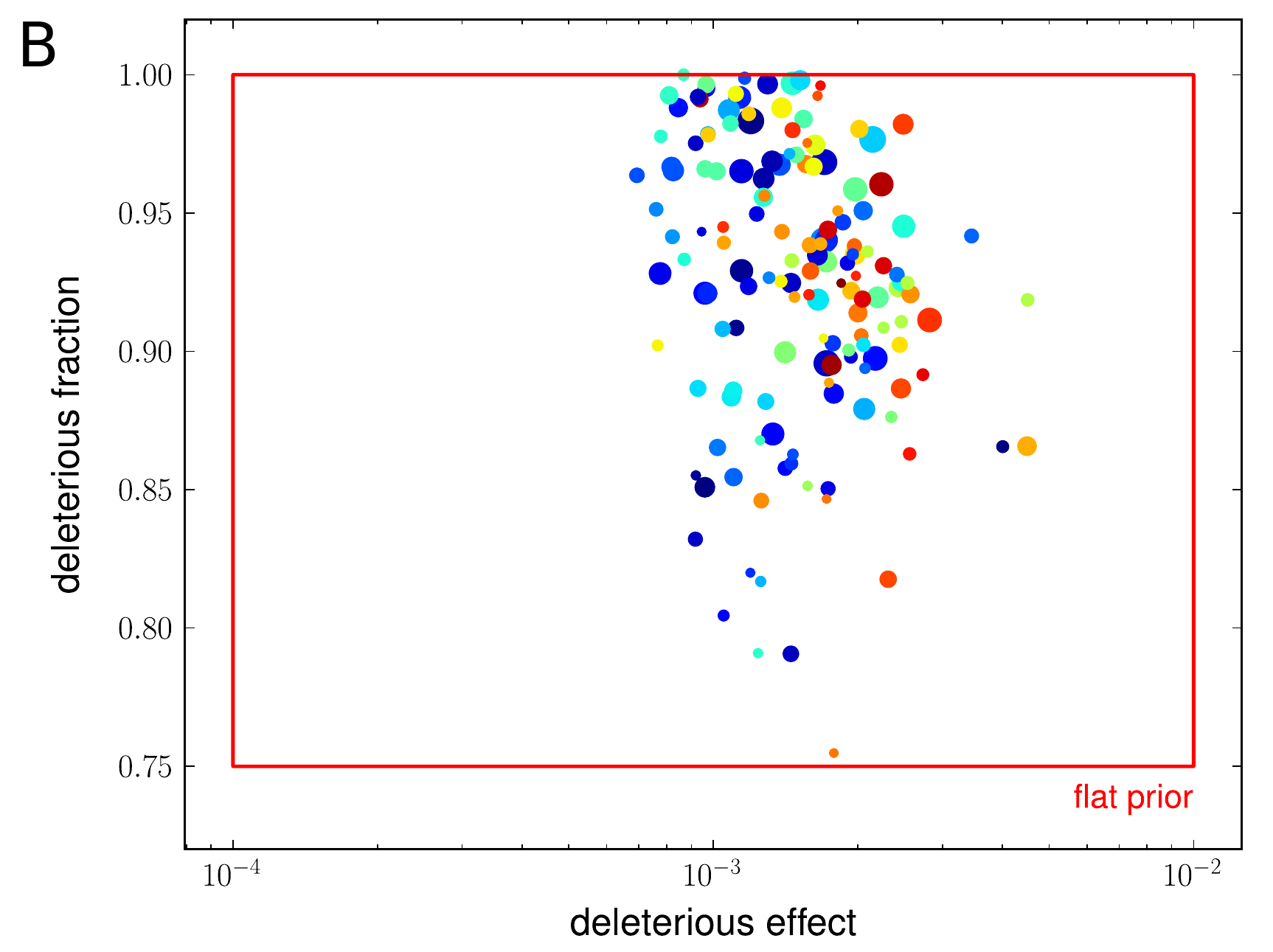}
\label{fig:simsfig}}
\caption{Distribution of selection coefficients on synonymous sites. Panel A)
The depression in $\pfix$ depends on the deleterious effect size 
of synonymous alleles. This parameter also reduces synonymous
diversity, measured by the probability of a derived allele to be found at
intermediate frequencies $P_\text{interm}$ (first inset).
Panel B) To assess the parameter space that affects synonymous fixation and
diversity, we run 2400 simulations with random parameters for deleterious effect
size, fraction of deleterious synonymous sites, average escape rate $\epsilon$
(color, blue to red corresponds to $10^{-2.5}$ to $10^{-1.5}$ per day), and rate of
introduction of new epitopes (marker size, from $10^{-3}$ to $10^{-2}$ per
day). Only simulations that reproduce the synonymous diversity and fixation
patterns observed in data are shown. These simulations demonstrate that
deleterious effects are around $-0.002$ and a large fraction of the 
synonymous mutations needs to be deleterious. As expected, larger
$s_d$ require larger $\epsilon$. Parameters are chosen
from prior distributions uniform in logspace as indicated by the red rectangle
(see methods).}
\label{fig:simheat}
\end{center}
\end{figure}

To get a better idea of the range of parameters that are compatible with the
observations and our interpretation, we performed computer simulations of
evolving viral populations assuming a mix of positive and purifying selection
and rare recombination.  For this purpose, we use the simulation package
FFPopSim, which includes a module dedicated to intrapatient HIV
evolution~\citep{zanini_ffpopsim:_2012}. For each simulation run, we specify the
deleterious effect of synonymous mutations, the fraction of synonymous mutations
that are deleterious, the escape rate (selection coefficient) of adaptive 
nonsynonymous mutations and the rate at which previously untargeted epitopes
become targeted (the latter determines the number of sites available for escape). 
Note that the escape rate is the sum of two
factors: (i) the beneficial effect due to the ability to evade the immune system
minus (ii) the fitness cost of the mutation in terms of structure, stability,
etc. Net escape rates in chronic infections have been estimated to be on the
order of $\epsilon = 0.01$ per day \citep{neher_recombination_2010,
Asquith:2006p28003}.

\FIG{simfixpvar} shows simulation results for the fixation probability and the
synonymous diversity for different deleterious effects of synonymous mutations.
We quantify synonymous diversity via $P_\text{interm}$, the fraction of sites
with an allele at frequency $0.25 < \nu < 0.75$. The synonymous diversity
observed in patient data is indicated in the figure.  To quantify the depression
of the fixation probability, we calculate the area between the measured fixation
probability and the diagonal, which is the neutral expectation
(\FIG{simfixpvar}, lower inset). If no fixation happens, the area will be
$-0.5$; if every mutation fixes, the area will be $+0.5$. In HIV-1 infected
patients, we find $P_\text{interm} \approx 0.005$, $A_\text{syn} \approx -0.2$
for synonymous changes and $A_\text{nonsyn} \approx 0$ for nonsynonymous
changes. In the three simulations shown in \FIG{simfixpvar}, the fixation
probability of synonymous alleles decreases from the neutral expectation
($A_\text{syn} \approx 0$) to zero ($A_\text{syn} \approx -0.5$) as their
fitness cost increases; the synonymous diversity plummets as well, as
deleterious mutations are selected against.

To map the parameter range of the model that is compatible with the data, we
repeatedly simulated the evolution with random choices for the parameters in
certain bounds, see \FIG{simsfig}. Among all simulations, we select the ones
that show $A_\text{syn}$ and $P_\text{interm}$ as observed in the data, i.e., a
large depression in fixation probability of synonymous mutations but,
simultaneously, a moderately high synonymous diversity. Specifically,
\FIG{simsfig} shows parameter combinations for which we found $A_\text{syn} <
-0.15$ and $0.0025 < P_\text{interm} < 0.010$. These conditions indicate that a
high fraction ($\gtrsim 0.8$) of sites has to be deleterious with effect size
$|s_d| \sim 0.002$.  This result fits well the expectation based on the
fixation/extinction times above (see \FIG{fixp1}). The results are plausible:
(i) a substantial depression in $\pfix$ requires pervasive deleterious
mutations, otherwise the majority of observed polymorphisms are neutral and no
depression is observed; (ii) in order to hitchhike, the deleterious effect size
has to be much smaller than the escape rate, otherwise the double mutant has
little or no fitness advantage. Consistent with this argument, larger
deleterious effects in \FIG{simsfig} correspond to larger escape rates; and (iii)
mutations with a deleterious effect smaller than approximately $0.001$ behave
neutrally, consistent with the typical coalescent times observed in HIV-1.

The above simulations show that hitchhiking can explain the observation of
deleterious mutations that rarely fix. However, in a simple model where
nonsynonymous escape mutations are unconditionally beneficial, they almost
always fix once they reach high frequencies -- $A_{\mathrm{nonsyn}}$ is well
above zero. This is incompatible with the blue line in \FIG{fixp2}: in an HIV-1
infection, nonsynonymous mutations at high frequency often disappear again, even
though many are at least transiently beneficial. Inspecting the trajectories of
nonsynonymous mutations suggests the rapid rise and fall of many alleles. We
test two possible mechanisms that are biologically plausible and could explain
the transient rise of nonsynonymous mutations: time-dependent selection and
within-epitope competition.

The former hypothesis can be formulated as follows: if the immune system
recognizes the escape mutant before its fixation, the mutant might cease to be
beneficial and disappear soon, despite its quick initial rise in frequency. In
support of this idea, \citet{richman_rapid_2003, bunnik_autologous_2008} report
antibody responses to escape mutants. These responses are delayed by a few
months, roughly matching the average time needed by an escape mutant to rise
from low to high frequency. To model this type of behavior, we assume that
antibody responses against escape mutations arise with a rate proportional to
the frequency of the escape mutation and abolish the benefit of the escape
mutations. As expected, this type of time-dependent selection retains the
potential for hitchhiking, but reduces fixation of nonsynonymous mutations.
\figurename~S\timedependence~shows that $\pfix$ of synonymous mutations is not
affected by this change, while $\pfix$ of nonsynonymous mutations approaches
the diagonal as the rate of recognition of escape mutants is increased. 

In the alternative hypothesis, several different escape mutations within the
same epitope might arise almost simultaneously and start to spread. Their
benefits are not additive, because each of them is essentially sufficient to
escape. As a consequence, several escape mutations rise to high frequency
rapidly, while the one with the smallest cost in terms of replication,
packaging, etc. is most likely to eventually fix. The emergence of multiple
sweeping nonsynonymous mutations in real HIV-1 infections has been shown
\citep{moore_limited_2009, bar_early_2012}. This scenario has been explicitly 
observed in  the evolution of resistance to 3TC, where the mutation 
M184V is often preceeded by M184I \citep{hedskog_dynamics_2010}. Similarly, AZT 
resistance often emerges via the competing TAM and TAM1 pathways.
Within epitope competition can be
implemented in the model through epistasis between escape mutations. While each
mutation is individually beneficial, combining the mutations is deleterious (no
extra benefit, but additional costs). Again, we find that the potential for
hitchhiking is little affected by within epitope competition but that the
fixation probability of nonsynonymous polymorphisms is reduced. With roughly six
mutations per epitope, the simulation data are compatible with observations; see
\figurename~S\withinepi. The two scenarios are not exclusive and possibly both
important in HIV-1 evolution.

\section{Discussion}
By analyzing the fate of mutations in longitudinal data of HIV-1 \env{} evolution,
we demonstrate selection against synonymous substitutions in the comparatively
conserved regions C2-C4 of the \env{} gene. Comparison with biochemical studies
of base pairing propensity in RNA genome of HIV-1 indicates that these
mutations are deleterious, at least in part, because they disrupt stems in RNA
secondary structures. Computational modeling shows that these mutations have
deleterious effects on the order of $0.002$ and that they are brought to high
frequency through linkage to adaptive mutations.

The fixation and extinction times and probabilities represent a rich and simple
summary statistics useful to characterize longitudinal sequence data and compare
to models via computer simulations. A method that is similar to ours {\it in
spiritu} has been recently used in a longitudinal study of influenza
evolution~\citep{strelkowa_clonal_2012}. The central quantity used in that
article, however, is a ratio between propagators of nonsynonymous and synonymous
mutations. The latter is used as an approximately neutral control; this method
can therefore not be used to investigate synonymous changes themselves. More
generally, evolutionary rates at synonymous sites are often used as a baseline
to detect purifying or diversifying selection at the protein level
\cite{Hurst:2002p32608}. It has been pointed out, however, that the rate of
evolution at synonymous sites varies considerable along the HIV-1 genome
\citep{mayrose_towards_2007} and that this variation can confound estimates of
selection on proteins substantially \citep{ngandu_extensive_2008}.

A functional significance of the insulating RNA structure stems between the
hypervariable loops has also been proposed previously
\citep{watts_architecture_2009, sanjuan_interplay_2011} and conserved RNA
structures exist in different parts of the HIV-1 genome. Since there are
of course many ways to build an RNA stem in a particular location, we do
not necessarily expect a strong signal of conservation in cross-sectional data. 
Our analysis, however, is able to
quantify the fitness effect of RNA structure within single infections and
demonstrates how selection at synonymous sites can alter genetic diversity and
dynamics. The observed hitchhiking highlights the importance of linkage due to
infrequent recombination for the evolution of HIV-1
\citep{neher_recombination_2010, batorsky_estimate_2011,
josefsson_majority_2011}. The recombination rate has been estimated to be on the
order of $\rho = 10^{-5}$ per base and day. It takes roughly $t_{sw} =
\epsilon^{-1} \log \nu_0$ generations for an escape mutation with escape rate
$\epsilon$ to rise from an initially low frequency $\nu_0\approx \mu$ to frequency
one. This implies that a region of length $l = (\rho t_{sw})^{-1} = \epsilon /
\rho \log \nu_0$ remains linked to the adaptive mutation. With $\epsilon=0.01$,
we have $l\approx 100$ bases. Hence we expect strong linkage between the
variable loops and the flanking sequences, but none far beyond the variable
regions, consistent with the lack of signal outside of \shankaregion. In case of
much stronger selection -- such as observed during early CTL escape or drug
resistance evolution -- the linked region is of course much larger
\citep{nijhuis_stochastic_1998}. 

While classical population genetics assumes that the dominant stochastic force
is genetic drift, i.e. non-heritable fluctuations in offspring number, our
results show that stochasticity due to linked selection is much more important.
Such fluctuations have been termed \emph{genetic draft} by
\citet{gillespie_genetic_2000}. Genetic draft in facultatively sexual population
such as HIV-1 has been characterized in \citep{neher_genetic_2011}. Importantly,
large population sizes are compatible with low diversity and fast coalescence
when draft dominates over drift.

Contrary to na\"ive expectations, the adaptive escape mutations do not seem to
be unconditionally beneficial. Otherwise we would observe almost sure fixation
of a nonsynonymous mutation once they reach intermediate frequencies. Instead, we
find that the fixation probability of nonsynonymous mutations is roughly given
by its frequency. There are several possible explanations for this observation.
Similar to synonymous mutations, the majority of nonsynonymous mutations could
be weakly deleterious, and the adaptive and deleterious parts could conspire to yield a
more neutral-like averaged fixation probability. While weakly deleterious 
nonsynonymous mutations certainly exist and will contribute to a depression of the
fixation probability, we have seen that a substantial depression requires that
weakly deleterious nonsynonymous polymorphisms at high frequency greatly 
outnumber escape mutations. This seems unlikely, since nonsynonymous diversity
exceeds synonymous diversity despite the overall much greater constraints on
the amino acid sequence. 

Alternatively, the lack of fixation could be due to time-dependent environment
through an immune system that is catching up, or competition between mutations
that mediate escape within the same epitope. We explore both of these
possibilities and find that both produce the desired effect in computer models. Furthermore, there
is experimental evidence in support of both of these hypotheses. Serum from HIV-1
infected individuals typically neutralizes the virus that dominated the
population a few (3-6) months earlier \citep{richman_rapid_2003}. This suggests that
escape mutations cease to be beneficial after a few months and might revert if
they come with a fitness cost. Deep sequencing of regions of \env{} after
antibody escape have revealed multiple escape mutations in the same epitope
\citep{moore_limited_2009, bar_early_2012}. Presumably, each one of these
mutations is sufficient for escape but most combinations of them do not provide
any additional benefit to the virus. Hence only one mutation will spread and the
others will be driven out of the population although they transiently reach high
frequencies. The rapid emergence of multiple escape mutations in the same
epitope implies a large effective population size that explores all necessary point
mutations rapidly. A similar point has been made recently by Boltz {\it et al.}
in the context of preexisting drug resistance mutations
\citep{boltz_ultrasensitive_2012}. 

Our results emphasize the inadequacy of independent site models of HIV-1 evolution
and the common assumption that selection is time independent or additive. 
If genetic variation is only transiently beneficial, existing estimates of the
strength of selection \citep{neher_recombination_2010,batorsky_estimate_2011}
could be substantial underestimates. Furthermore, weak conservation and
time-dependent selection result in estimates of evolutionary 
rates that depend on the time interval of observation, with lower rates across
larger intervals. This implies that deep nodes in phylogenies might be older than 
they appear.

\section{Methods}
\subsection{Sequence data collection}
Longitudinal intrapatient viral RNA sequences were collected from published
studies \citep{shankarappa_consistent_1999, liu_selection_2006,
bunnik_autologous_2008} and downloaded from the Los Alamos National Laboratory
(LANL) HIV sequence database~\citep{LANL2012}. The samples from some patients
show substantial population structure and were discarded (see
\figurename~S\PCApat); a total of 11 patients with 4-23 time points each and
approximately 10 sequences per time point were analyzed. The time intervals
between two consecutive sequences ranged from 1 to 34 months, most of them
between 6 and 10 months.

\subsection{Sequence analysis}
The sequences were translated and the resulting amino acid sequences aligned
using Muscle~\citep{edgar_muscle:_2004} to each other and the NL4-3 reference
sequences separately for each patient. Within each patient, the consensus
nucleotide sequence at the first time point was used to classify alleles as
``ancestral'' or ``derived'' at all sites. Sites that include large
frequencies of gaps were excluded from the analysis to avoid artifactual
substitutions due to alignment errors. Allele frequencies at different time
points were extracted from the multiple sequence alignment.

A mutation was considered synonymous if it did not change the amino acid
corresponding to the codon, and if the rest of the codon was in the ancestral
state. Codons with more than one mutation were discarded. Slightly different
criteria for synonymous/nonsynonymous discrimination yielded similar results.

\subsection{Fixation probability and secondary structure}
For the estimates of time to fixation/extinction, polymorphisms were binned by
frequency and the time to first reaching either fixation or extinction was
stored. The fixation probability was determined as the long-time limit of the
resulting curves. Mutations that reached high frequency but neither fixed nor
were lost were classified as ``floating'', with one exception: if they first
reached high frequencies within 3 years of the last time point, it was assumed
they had not had sufficient time to settle, so they were discarded.

The SHAPE scores quantifying the degree of base pairing of individuals sites in
the HIV-1 genome were downloaded from the journal website
\citep{watts_architecture_2009}. Wherever possible, SHAPE reactivities were
assigned to sites in the multiple sequence alignments for each patient through
the alignment to the sequence of the NL4.3 virus used in
ref.~\citep{watts_architecture_2009}. Problematic assignments in indel-rich
regions were excluded from the analysis. The variable loops and flanking
regions were identified manually starting from the annotated reference HXB2
sequence from the LANL HIV database~\citep{LANL2012}. 

\subsection{Computer simulations}
Computer simulations were performed using FFPopSim
\citep{zanini_ffpopsim:_2012}. Briefly, FFPopSim enables individual-based
simulations where each site in the genome is represented by one bit that can be
in one of two states. Outcrossing rates, crossover rates, mutations rates and
arbitrary fitness functions can be specified. We used a generation time of 1
day, an outcrossing rate of $r=0.01$ per day \citep{batorsky_estimate_2011,
neher_recombination_2010}, a mutation rate of $\mu=10^{-5}$
\citep{mansky_lower_1995, abram_nature_2010} and simulated intrapatient
evolution for 6000 days. For simplicity, third positions of every codon were
deemed synonymous and assigned either a selection coefficient $0$ with
probability $1-\alpha$ or a deleterious effect $s_d$ with probability $\alpha$.
Mutations at the first and second positions were assigned strongly deleterious 
fitness effects 0.02. At 
rate $k_A$, a random locus in the genome is designated an epitope that can
escape by one or several mutations with an exponentially distributed escape rate
with mean $\epsilon$. Both full-length HIV-1 genomes and \env{}-only simulations
were performed and yielded comparable results.

The simulations were repeated 2400 times with random choices for the following
parameters: the fraction of deleterious sites $\alpha$ was sampled uniformly
between 0.75 and 1.0; the average deleterious effect $s_d$ was sampled such that
its logarithm was uniformly distributed  between $10^{-4}$ and $10^{-2}$; the
average escape rate $\epsilon$ of escape mutation was sampled such that its logarithm was
uniform between $10^{-2.5}$ and $10^{-1.5}$ and the rate $k_A$ of new antibody
challenges such that its logarithm was uniform between $10^{-3}$ and $10^{-2}$
per generation. Populations were initialized with a homogenous founder
population and were kept at an average size of $N=10^4$ throughout the
simulation. After 30 generations of burn-in to create genetic diversity, new
epitopes were introduced at a constant rate $k_A$. 

For the models with competition within epitopes, a complex epistatic fitness
landscape was designed such that each single mutant is sufficient for full
escape. In particular, each mutation had a linear effect equal to the escape,
but a negative epistatic effect of the same magnitude between each pair of sites
was included. Higher order terms compensated each other to make sure that not
only double mutants, but all k-mutants with $k \geq 1$ had the same fitness (see
supplementary materials). To model recognition of escape variants by the immune
system  catching up, the beneficial effect of an escape mutation was set
to its previous cost of -0.02 with a probability per generation proportional to
the frequency of the escape variant.

For each set of parameters, fixation probabilities and probabilities of
synonymous polymorphisms $P_\text{interm}$ were calculated as averages over
100 repetitions (with different random seeds).

The areas below or above the neutral fixation probability (diagonal line) were
estimated from the binned fixation probabilities using linear interpolation
between the bin centers. This measure is sufficiently precise for our purposes.
In 10 runs out of 2400, the highest frequency bin was empty so the fixation
probability could not be calculated; those runs were excluded from
\FIG{simsfig}.

\subsection{Methods availability}
All analysis and computer simulation scripts, as well as the sequence alignments
used, are available for download at \url{http://git.tuebingen.mpg.de/synmut}.

\section*{Acknowledgements}
We thank Jan Albert, Trevor Bedford and Pleuni Pennings and members of the lab for 
stimulating discussions and critical reading of the manuscript.
This work is supported by the ERC starting grant HIVEVO 260686 and 
in part by the National Science Foundation under Grant No.~NSF PHY11-25915.

\bibliographystyle{natbib}

\newpage
\appendix
\onecolumngrid
\setcounter{figure}{0}
\include{supplement_body}
\end{document}

%% file: supplement_body.tex

\makeatletter 
\renewcommand{\thefigure}{S\@arabic\c@figure}
\makeatother

\section{Selection of the patient data}
\begin{figure}[ht]
\begin{center}
\includegraphics[width=0.35\linewidth]{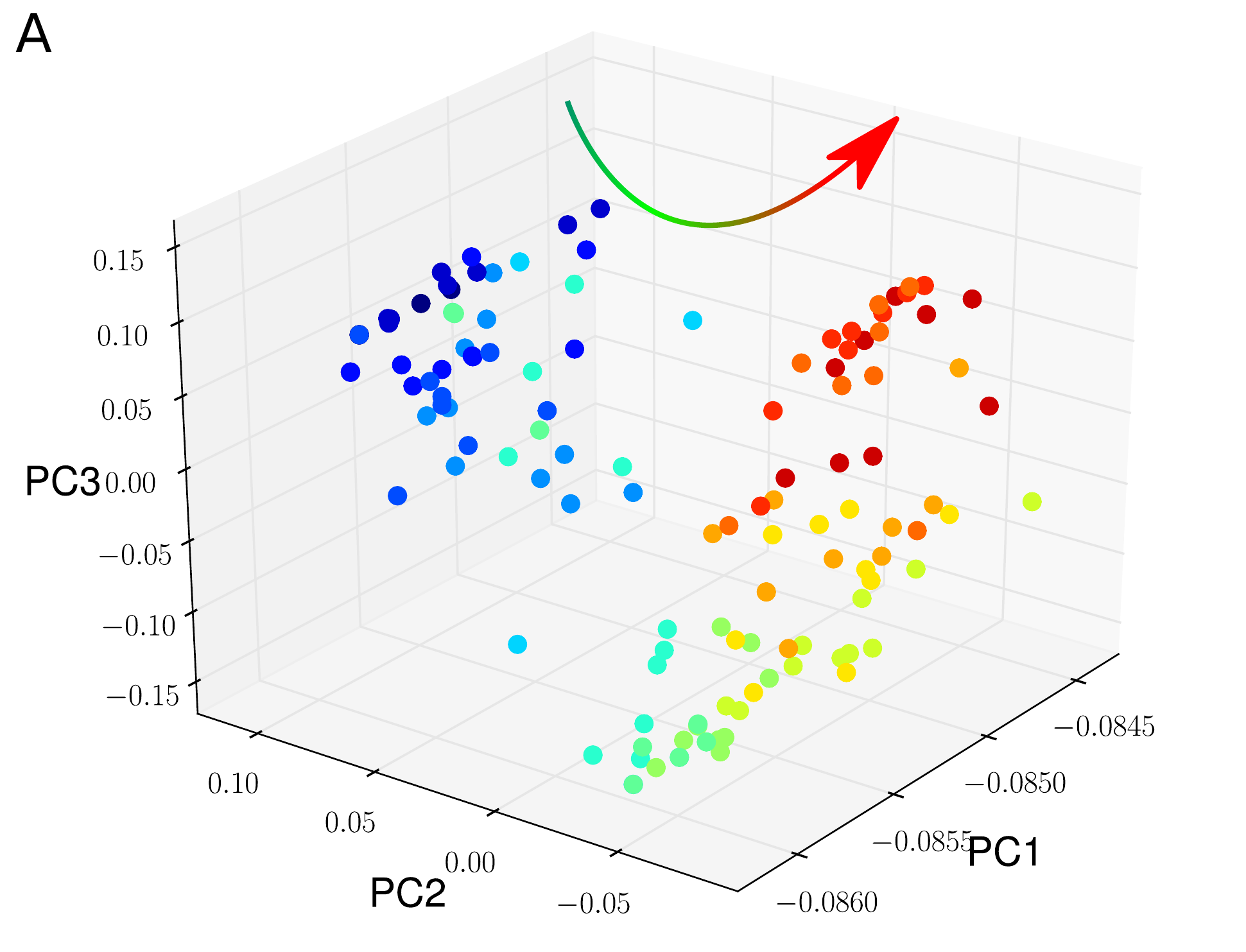}
\includegraphics[width=0.35\linewidth]{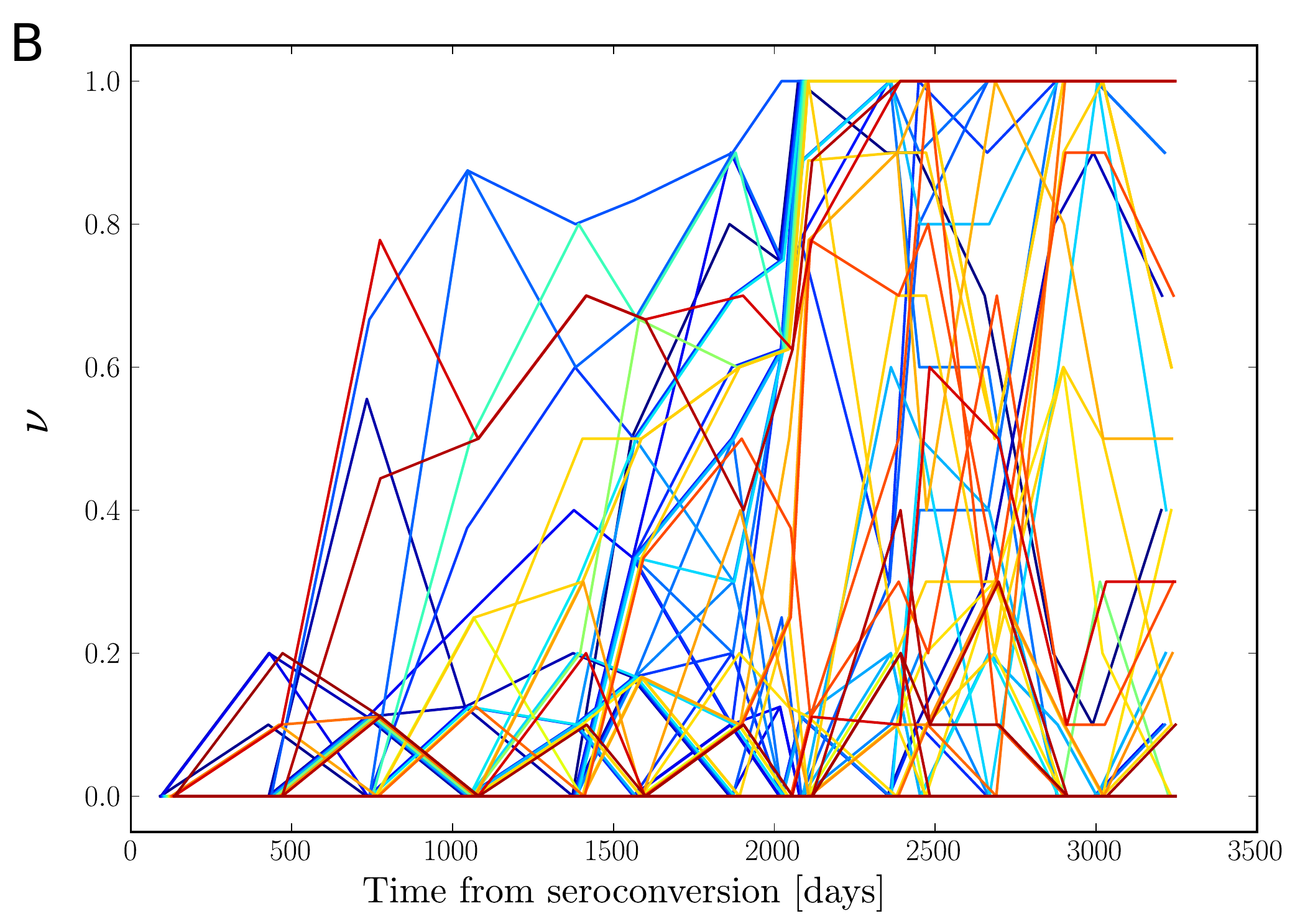}
\includegraphics[width=0.35\linewidth]{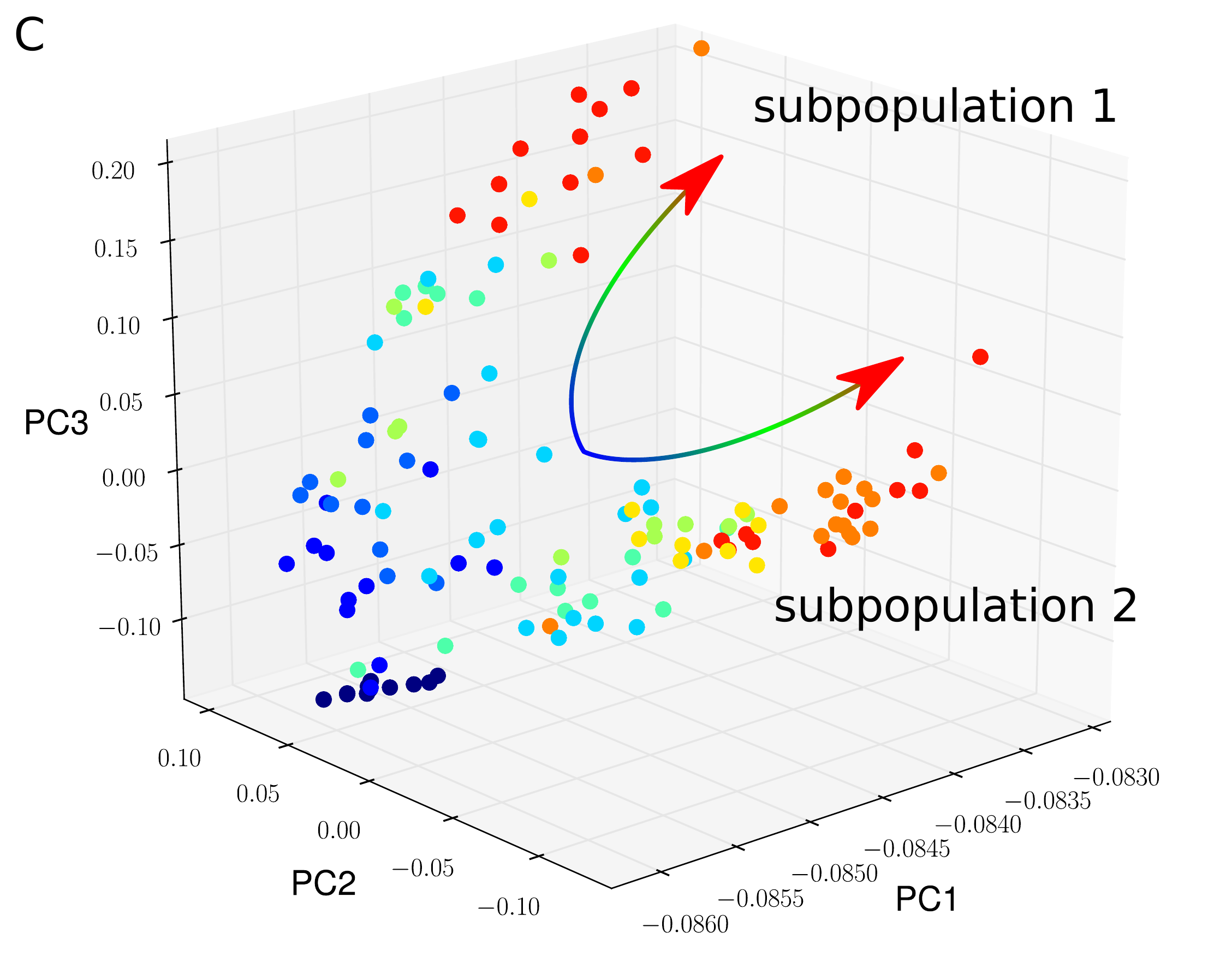}
\includegraphics[width=0.35\linewidth]{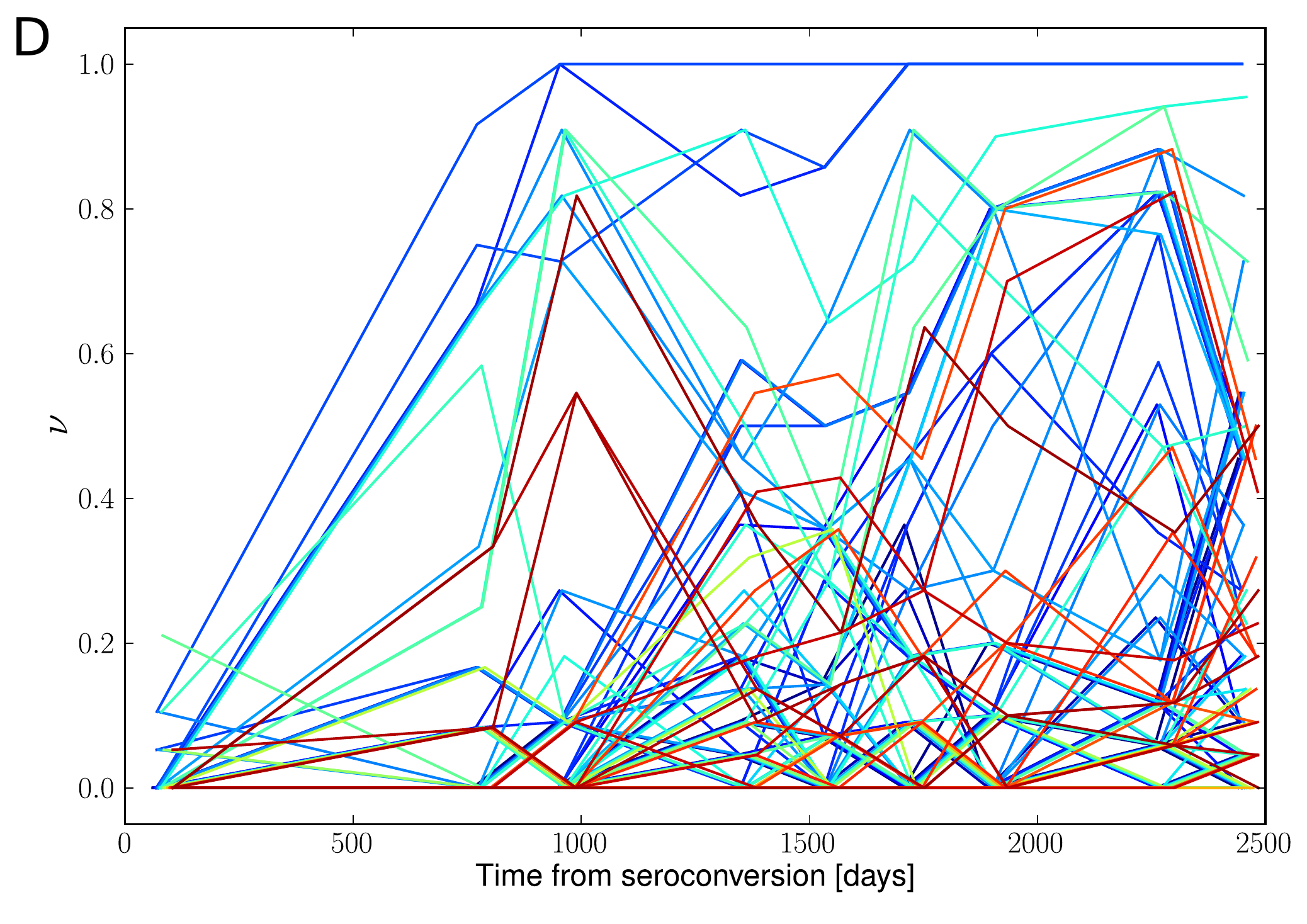}
\caption{Structure of viral populations and patient selection.
Panel A) shows a PCA of all sequences from patient p1 (colors indicate time from
seroconversion, from blue to red). Panel B) shows allele frequency trajectories for nonsynonymous
changes in the same patient. Here, the blue to red color map corresponds to the
position of the allele in \env{} from 5' to 3'. Panels C) and D) show analogous
plots for data from patient p7. Samples after day 1000 split into two clusters in the PCA and no mutations that arise after day 1000 fix, presumably because they are restricted
to one subpopulation. All patients like p7 (p4, p7, p8, p9 from ref.~\citealp{shankarappa_consistent_1999} and
ACH19542 and ACH19768 from ref.~\citealp{bunnik_autologous_2008}) were excluded
from our analysis.}
\label{fig:aftp}
\end{center}
\end{figure}

\newpage
\section{Synonymous diversity across the HIV genome}
\begin{figure}[h]
\begin{center}
\includegraphics[width=\linewidth]{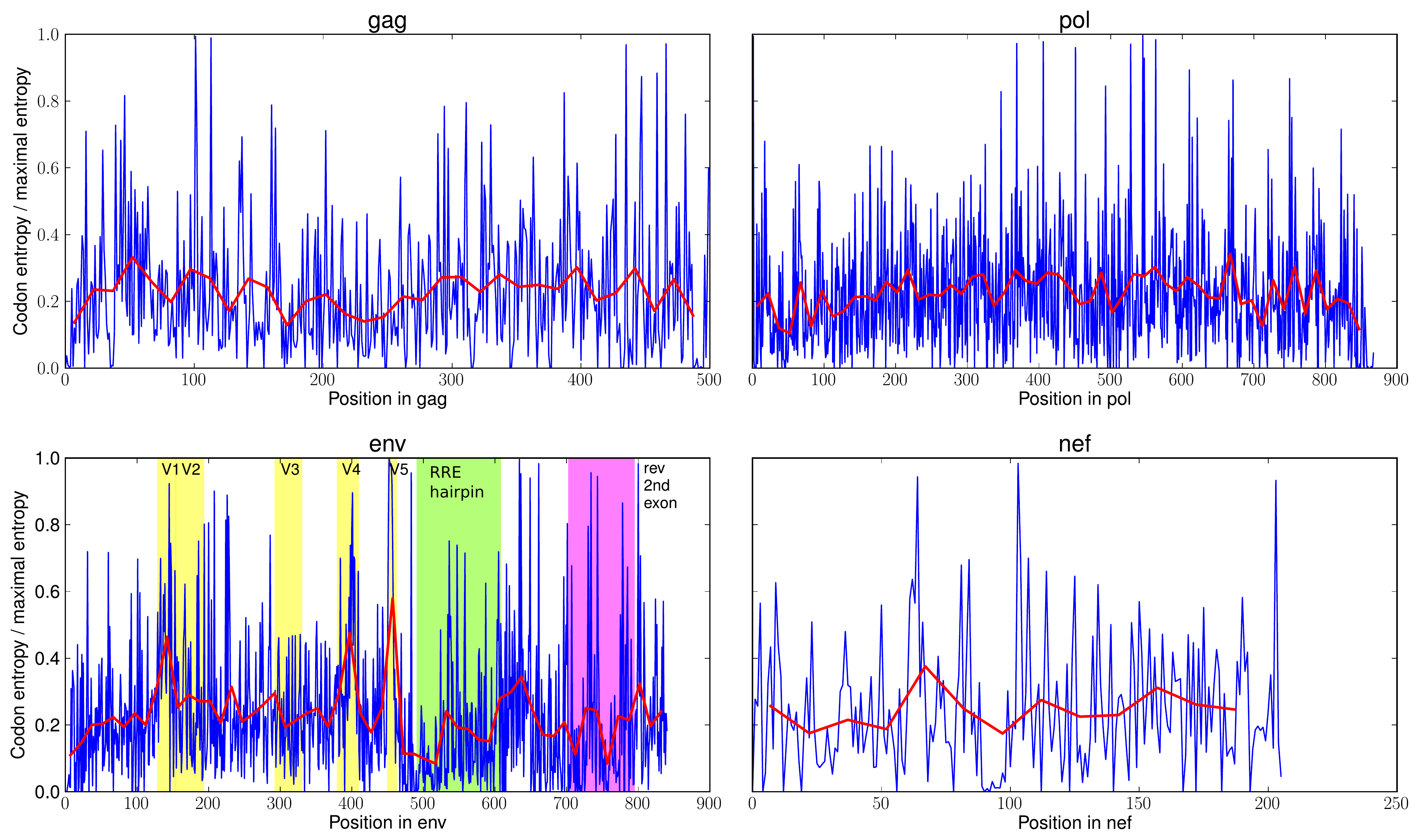}
\caption{
Synonymous diversity across the HIV genome, as quantified by the normalized
codon entropy among sequences coding for the consensus amino acid. In most
parts of the genome, synonymous sites show little conservation. The synonymous
diversity peaks at the variable regions in {\it env} and is reduced in regions 
under purifying selection (RRE hairpin, second {\it tat}/{\it rev} exons). The
normalized codon entropy is calculated as follows (see the script
\texttt{codon\_entropy\_synonymous\_subtypeB.py} for the full algorithm): (i)
from a subtype B multiple sequence alignment (MSA) from the LANL website (filtered sequences only, version 2011)~\cite{LANL2012}, we calculate the
consensus amino acid at each position in the HIV genome; (ii) we count how often
each codon coding for the consensus amino acid appears in the MSA; (iii) at each
amino acid position, we divide by the number of sequences in the MSA that had
the consensus amino acid at that position, obtaining {\it codon frequencies}
$\nu_c$; (iv) we calculate the codon entropy from each position as: $S := -
\sum_{c} \nu_c \log \nu_c$, where $c$ runs over codons that code for the
consensus amino acid at this site; (v) we divide by the maximal codon entropy of
that amino acid (e.g. $\log 2$ for twofold degenerate codons). All parts of
{\it env} that are part of a different gene (signaling peptide, second {\it rev}
exon) have been excluded from our main analysis, to avoid contamination by
protein selection in a different reading frame.
Note that all gap-rich columns of the MSA are stripped from this figure, so genes such as {\it env} might appear shorter than they actually
are.
}
\label{fig:syndiv_genome}
\end{center}
\end{figure}
\newpage
%

\section{Time-dependent selection}
\begin{figure}[h]
\begin{center}
\includegraphics[width=0.6\linewidth]{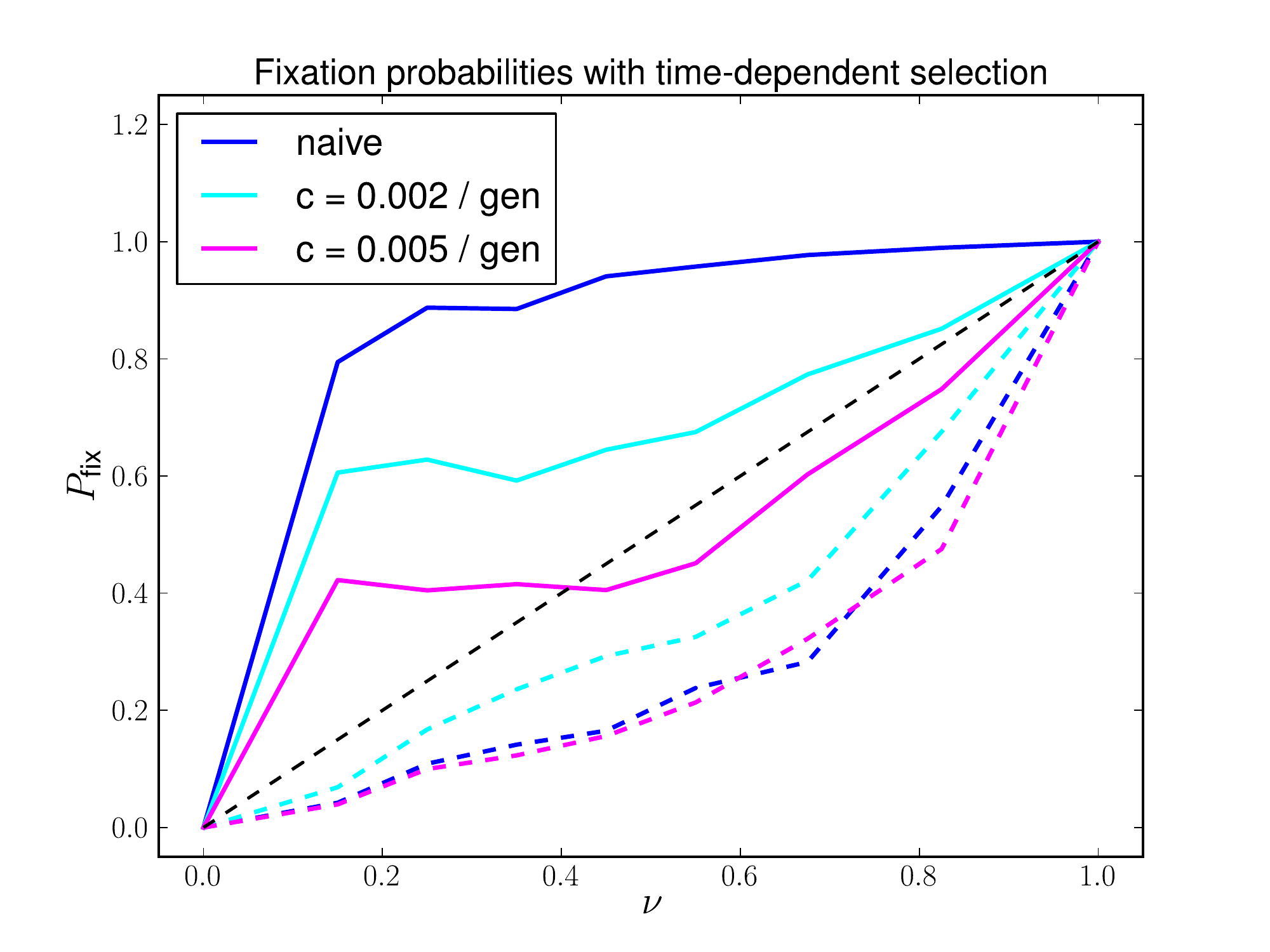}
\caption{
Time-dependent selection reduces fixation of nonsynonymous mutations. The figure
compares the fixation probability in the time independent model (na\"ive) to
a model with time dependent selection that mimics  an evolving immune system.
It has been found that virus is typically neutralized by serum from a few months
earlier~\citep{richman_rapid_2003} but not by contemporary serum. We model this
evolving immune system by assuming that escaped variants lose their beneficial
effect with a rate proportional to the frequency of the escaped variant. 
Specifically, the selection effect of the escape mutations is
reset to its fitness cost of $-0.02$ with probability
\[ P_\text{recognized}(t) = c \cdot \nu(t), \] 
per generation, where $c$ is a constant coefficient shown in the legend that
encodes the overall efficiency of the host immune system. With increasing
probability of recognition, the fixation of frequent escape mutants is reduced,
while hitch-hiking of synonymous mutations is not affected. The precise
shape of $\pfix(\nu)$ depends on the details of the $P_\text{recognized}(t)$, and 
we do not think that the high $\pfix(\nu)$ for $\nu<0.2$ is meaningful.
The other parameters for the shown simulations are
the following: deleterious effect $s_d = 10^{-3}$, average escape rate $\epsilon = 0.016$,
fraction of deleterious synonymous mutations $\alpha = 0.986$, rate of new epitopes
$k_A=0.0014$ per generation.
}
\label{fig:tds}
\end{center}
\end{figure}

\newpage
\section{Within-epitope competition}
\begin{figure}[h]
\begin{center}
\includegraphics[width=0.6\linewidth]{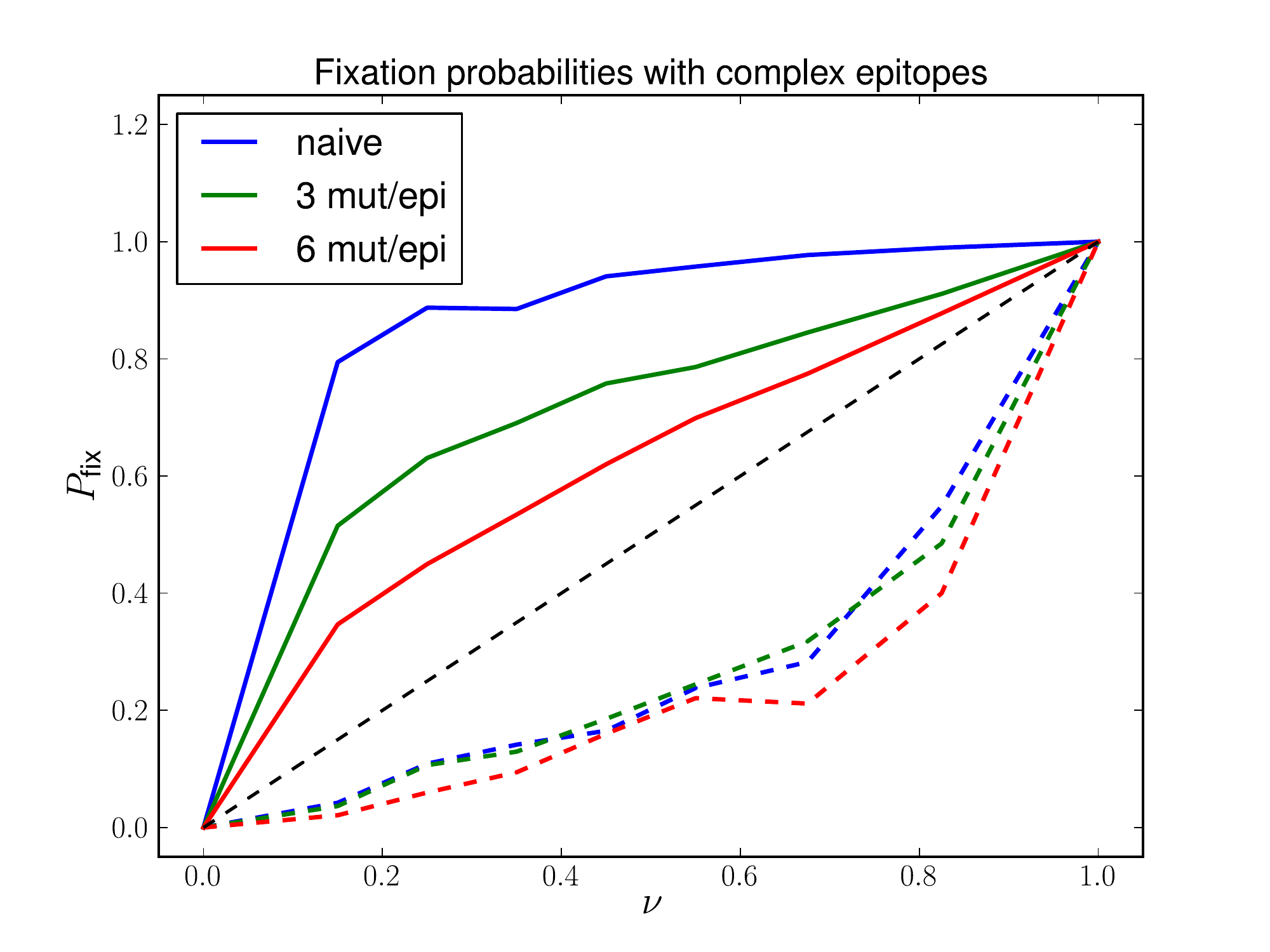}
\caption{
Competition between escape mutations in the same epitope reduces fixation of
nonsynonymous mutations. The figure compares the fixation probability of models
with one, three, or six mutually exclusive escape mutations within the
same epitope. Within epitope competition results in reduced fixation
probabilities of nonsynonymous changes, whereas the synonymous changes behave 
similarly in all cases. We assume that escape can happen at $n$ sites out of 3
consecutive codons and vary $n$.
The fitness landscape of each epitope includes negative epistatic terms, so that
the joint presence of more than one escape mutation is not any more beneficial
for the virus than a single mutation. Specifically, each site has two alleles,
$\pm 1$, where $-1$ is the ancestral one and $+1$ the derived one; the fitness
coefficient of a $k$-tuple of sites within the epitope is $f_k = (-1)^{k-1}
2^{1-n}\eta_\epsilon $, where $\eta_\epsilon$ is the escape rate of the epitope
drawn from an exponential distribution with mean $\epsilon$ and 
$n$ is the number of competing escapes in the epitope. 
In this evolutionary scenario, many escape mutations start to sweep on different backgrounds within the viral population, but eventually
compete and only one of them fixes. The other parameters for the shown simulations are
the following: deleterious effect $s_d = 10^{-3}$, average escape rate $\epsilon = 0.016$,
fraction of deleterious synonymous mutations $\alpha = 0.986$, rate of new epitopes
$k_A=0.0014$ per generation.
}
\label{fig:wec}
\end{center}
\end{figure}